\documentclass[5p]{elsarticle}

\usepackage{lineno,hyperref}
\usepackage{gensymb}
\usepackage[T1]{fontenc}
\usepackage{graphicx}
\usepackage{subcaption}
\usepackage{url,array}
\usepackage{color,soul,xcolor}
\usepackage{textcomp}
\usepackage{longtable}
\usepackage{graphicx}
\usepackage{dcolumn}
\usepackage{bm}
\usepackage{color,soul,xcolor,amsmath}
\usepackage{amssymb}
\usepackage{amsthm}

\graphicspath{ {./images/} }
\modulolinenumbers[5]

\newcommand{\beq}{\begin{equation}}
\newcommand{\eeq}{\end{equation}}
\newcommand{\bdis}{\begin{displaymath}}
\newcommand{\edis}{\end{displaymath}}
\newcommand{\bea}{\begin{eqnarray}}
\newcommand{\eea}{\end{eqnarray}}
\newcommand{\barr}{\begin{array}}
\newcommand{\earr}{\end{array}}
\newcommand{\bfig}{\begin{figure}[!]}
\newcommand{\efig}{\end{figure}}

\begin{document}

\begin{frontmatter}

\title{Magnetocaloric effect at the reorientation of the magnetization in ferromagnetic multilayers with perpendicular anisotropy}

\author[INRIM]{Vittorio Basso, Carlo P. Sasso}
\author[CNRS]{Martino LoBue}
\author[brooklyn,gradcenter]{Karl G. Sandeman\corref{mycorrespondingauthor}}
\cortext[mycorrespondingauthor]{Corresponding author}
\ead{karlsandeman@brooklyn.cuny.edu}

\affiliation[INRIM]{Istituto Nazionale di Ricerca Metrologica, Strada delle Cacce 91, 10135 Torino, Italy}
\affiliation[CNRS]{Universit\'e Paris-Saclay, ENS Paris-Saclay, CNRS, SATIE, 91190, Gif-sur-Yvette, France}
\affiliation[brooklyn]{Department of Physics, Brooklyn College of The City University of New York, 2900 Bedford Avenue, Brooklyn, NY 11210 USA}
\affiliation[gradcenter]{Physics Program, The Graduate Center, CUNY, 365 Fifth Avenue, NY 10016, USA}

\begin{abstract}
We investigate the magnetocaloric effect obtained by the rotation of a magnetic field applied to an exchange-coupled multilayer system composed of two different ferromagnetic (FM) materials. We specifically consider a system in which the two FMs have perpendicular uniaxial anisotropy axes and utilise conditions which yield a reorientation of the total magnetization when compensation between the anisotropies of the two layers occurs. We calculate the consequent entropy change associated with the ``artificial" reorientation. By using known parameters from MnBi and Co we predict an entropy change of $\Delta s = 0.34$ Jkg$^{-1}$K$^{-1}$ for perfect coupling. Lastly, we study the behavior of the multilayer under a rotating magnetic field via a micromagnetic model. When the layer thicknesses are of the order of the local domain wall width, the magnetic field-induced entropy change can be obtained with magnetic fields one order of magnitude lower than in the uncoupled case. 
\end{abstract}


\end{frontmatter}

\section{Introduction}
The magnetocaloric effect (MCE) is the thermodynamic response of a material to a change in the level of an applied magnetic field.  It is quantified in terms of an isothermal entropy change, $\Delta s$ and an adiabatic temperature change, $\Delta T_{\rm ad}$ and was first observed by Weiss and Piccard in 1917~\cite{Weiss-1917}.  A resurgence of interest in the measurement and application of the MCE has resulted from the more recent discovery of giant MCEs in a range of materials with first order magnetic phase transitions~\cite{Sandeman-2012}.  In almost all of those cases, the MCE is associated with the ordering or disordering of magnetic spins.  

However, the magnetocaloric effect (MCE) also has an additional contribution from the magnetic anisotropy energy \cite{Kuzmin-1991}. Even if the entropy change is not high with respect to other magnetocaloric materials \cite{Kuzmin-2007}, this fact creates an interesting possibility for magnetocaloric cooling since the thermal effects are generated by simply turning a magnetic field with constant amplitude \cite{Kuzmin-1991}. The magnetocaloric effect associated with a spin reorientation transition has been observed in hexagonal ferrites \cite{Naiden-1997, Naiden-1997b, Basso-2011b, LoBue-2012} and in compounds with rare earths such as Er$_2$Fe$_{14}$B \cite{Skokov-2009, Basso-2011} and NdCo$_5$ \cite{Nikitin-2010}.   In this paper we investigate the possibility of creating an artificial spin reorientation by exchange coupling of two ferromagnetic materials with uniaxial anisotropies oriented in perpendicular directions. 

For uniaxial anisotropy, the Gibbs free energy density of a magnetic material can be expressed as:

\beq
\begin{aligned}
g_L(T, H; \theta) = & K_1(T) \sin^2\theta \\
& - \mu_0 M_s(T) \left( H_{\|}\cos\theta + H_{\bot} \sin\theta\right)
\end{aligned}
\eeq

\noindent where $K_1$ is the anisotropy constant, $\theta$ is the angle formed by the magnetization vector with the anisotropy axis, $M_s$ is the saturation magnetization and $H_{\|}$ and $H_{\bot}$ are the component of the field parallel and perpendicular to the magnetization. The entropy is given by the expression

\beq
s = - \frac{\partial g_L(H,T;\theta_{eq})}{\partial T}
\eeq

\noindent where $\theta_{eq}$ is given by the minimization of the energy by the conditions $\partial g_L/\partial \theta =0$ and $\partial^2 g_L/\partial \theta^2 > 0$. The entropy change is then 

\beq
\Delta s = -\frac{dK_1}{dT} \sin^2\theta_{eq} + \mu_0 \frac{dM_s}{dT} \left( H_{\|}\cos\theta_{eq} + H_{\bot} \sin\theta_{eq}\right) 
\eeq

\noindent The term proportional to $dM_s/dT$ describes the contribution associated with ordering or disordering of magnetic spins, while the term proportional to $dK_1/dT$ is an additional contribution related to the magnetic anisotropy energy. The microscopic origin of the term proportional to $dK_1/dT$ has to be found in the anisotropic magnetization mechanism \cite{Callen-1960}: the spin system is more disordered (high entropy state) when the magnetization points along an hard direction. A particularly interesting case is when $K_1$ is positive and the magnetic field is applied parallel (${\|}$) or perpendicular (${\bot}$) to the easy axis \cite{Basso-2011}. We have for (${\|}$):

\beq
\Delta s (H_{\|}) =  \mu_0 \frac{dM_s}{dT} H_{\|}
\eeq

\noindent and for (${\bot}$)

\beq
\Delta s (H_{\bot}) = -\left(\frac{dK_1}{dT} - \mu_0 \frac{dM_s}{dT} H_{AN} \right)\left(\frac{H_{\bot}}{H_{AN}}\right)^2
\eeq

\noindent for $H_{\bot}<H_{AN}$ and

\beq
\Delta s (H_{\bot}) = - \frac{dK_1}{dT} + \mu_0 \frac{dM_s}{dT} H_{\bot}
\eeq

\noindent for $H_{\bot}>H_{AN}$ where $H_{AN}$ is the anisotropy field $H_{AN}=2 K_1/\mu_0 M_s$. At $H_{\bot}<H_{AN}$ the entropy change depends on the difference ${dK_1}/{dT} - \mu_0 ({dM_s}/{dT}) H_{AN}$. If the temperature derivatives of the saturation magnetization, $dM_s/dT$, and of the anisotropy constant, $dK_1/dT$, both have the same sign, the resulting MCE under alternating field will be the difference of the two contributions giving no possibility to exploit the anisotropy effect. However, at magnetic fields above the anisotropy field, and for a magnetic field rotation from $(\bot)$ to $(\|)$ the term $dM_s/dT$ is eliminated and only the anisotropy contribution remains:

\beq
\Delta s (H_{\|}) - \Delta s (H_{\bot}) = \frac{dK_1}{dT}
\eeq

\noindent  When looking for relevant materials, one has to consider that the deviation from the easy axis must be obtained by not too large fields, (i.e. small with respect to $H_{AN}$). This situation naturally occurs in materials with a spin reorientation transition in which the main anisotropy constant $K_1$ crosses zero at the spin reorientation temperature, $T=T_{SR}$. As we describe in the next section, our model of an artifical SRT in a ferromagnetic multialyer requires us to find conditions on the anisotropy properties of the two layers such that the overall anisotropy of the multilayer vanishes at a particular temperature.

The remainder of this article is organised as follows: in section~\ref{A-SRT} we describe our model of a system of two, magnetically hard ferromagnetic layers with perpendicular anisotropy axes, and make a prediction of the magnitude of the rotating field MCE if MnBi and Co are used as the two layer materials.  Results of the micromagnetic model of exchange coupling are given in section~\ref{results} and conclusions are drawn in section~\ref{conclusions}.

\section{Artificial reorientation model}
\label{A-SRT}

The field of thin film spin reorientation (SRT) has been explored widely, albeit not  with a magnetocaloric motivation.  SRT studies have included those of ultrathin ferromagnetic films, where film thickness, temperature or magnetic field can be used to drive a transition from out-of-plane to in-plane magnetisation~\cite{Jensen-2006}.  Theoretical work predicted that a ferroelectric layer be used to provide electrical switching of the magnetisation of a neigbouring ferromagnetic layer, via SRT~\cite{Pertsev-2008} and this concept has been demonstrated in Cu/Ni multilayers on BaTiO$_3$~\cite{Shirahata-2015}.

Our focus is on the magnetocaloric properties of a layered system in which the SRT is of a different physical origin to that found in ultrathin ferromagnetic films.  We consider a multilayer stacked in the $x$-direction in which two ferromagnetic materials A and B both have uniaxial anisotropy but perpendicular orientations of their easy axes in the $y-z$ plane. As a specific case we consider material A with an easy axis directed along $z$ and material B with an easy axis directed along $y$ perpendicular to $z$. Considering uniaxial anisotropy to the first order, we have that the anisotropy energy of A is $K_{1_A} \sin^2\theta_A$ while for B is $K_{1_B} \sin^2(\theta_B-\pi/2)$ where $\theta_A$ and $\theta_B$ are the angles between the magnetisation of the A- or B-layer and the easy axis of the A-layer. For independent layers (without interlayer exchange) the free energy density is the sum of the two individual layer energies:

\beq
f_L = v_A K_{1_A} \sin^2\theta_A + v_B K_{1_B} \sin^2(\theta_B-\pi/2) \, ,
\eeq

\noindent where $v_A$ and $v_B$ are the volume fractions of A and B. As the anisotropy of B can be expressed as $K_{1_B}\sin^2(\theta_B-\pi/2)=K_{1_B}\cos^2\theta_B=K_{1_B} (1-\sin^2\theta_B)$, the orientation of the easy axis of layer B along $y$ corresponds to an effective negative anisotropy constant (along $z$). We may then use energies $K_{A} \sin^2\theta$ with $K_{A}=K_{1_A}>0$ for A and $K_{B} \sin^2\theta$ with $K_{B}=-K_{1_B}$ and $K_{B}<0$ for B. 

When interlayer exchange is taken into account one has to make a full micromagnetic model of the multilayer. If the exchange interaction dominates, the magnetization of A and B will be perfectly coupled $\theta=\theta_A=\theta_B$. The free energy will be:

\beq
f_L = \left(v_A K_A(T)+v_B K_B(T)\right)\sin^2\theta \, .
\eeq

\begin{figure}[htb]
\centering
\includegraphics[width=\columnwidth]{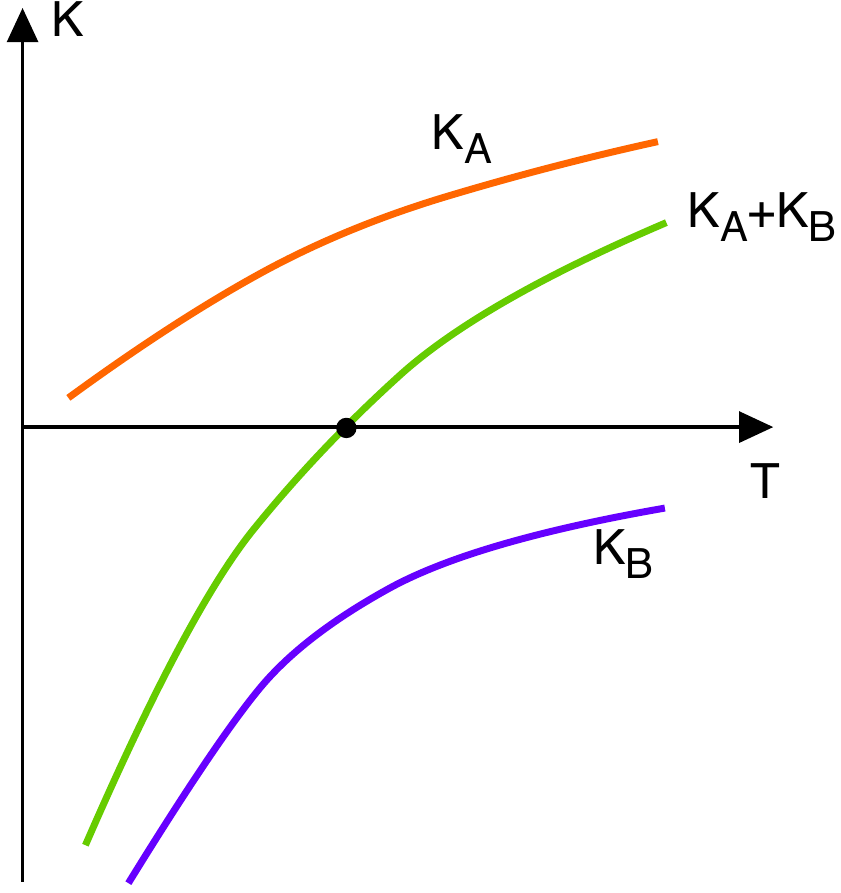}
\caption{Sketch of the anisotropy constants as a function of temperature for layers A and B which give a maximum entropy change at the transition temperature ($K_A+K_B=0$).} 
\label{FIG:0}
\end{figure}

\noindent If the condition $v_AK_A = -v_BK_B$ is realized the total anisotropy will vanish. If the anisotropy constants are both temperature dependent, there will be a particular temperature $T_{SR}$ at which this condition will be fulfilled and at which we will have an \emph{artificial} reorientation of the magnetization. The contribution to the entropy is $s=-df/dT$ and the entropy difference between the state of the multilayer in a magnetic field  which is parallel $(\|)$ or perpendicular $(\bot)$ with respect to $z$ will be $\Delta s = s_{\|}-s_{\bot}$:

\beq
\Delta s = v_A \frac{dK_A}{dT}+v_B \frac{dK_B}{dT} \, ,
\label{eq:s}
\eeq

\noindent because with $\theta=0$ $(\|)$ $s_{\|}=0$ and with $\theta =\pi/2$ $(\bot)$ $s_{\bot}=- (v_A {dK_A}/{dT}+v_B {dK_B}/{dT})$. If the A material has $dK_A/dT>0$ and the B material has $dK_{B}/dT>0$ (i.e. $dK_{1_B}/dT<0$) the two contributions in Equation~\ref{eq:s} will add constructively as shown in Figure~\ref{FIG:0} and the entropy change between $(\|)$ and $(\bot)$ will be maximized.
   
\subsection{Predictions of the properties of a composite Co and MnBi system}

To maximize the MCE under field rotation we select two materials A and B such that $dK_{1_A}/dT>0$ and $dK_{1_B}/dT<0$. As an example we take MnBi as material A and Co as material B. The selected materials have the following intrinsic properties at room temperature ($T = 293$ K).
 
\begin{itemize}
\item The low temperature phase (LTP) of MnBi has $\mu_0M_s=0.72$ T, $K_1=9\times10^{5}$ Jm$^{-3}$ \cite{Coey-2010}, $dK_1/dT=5.5\times10^{3}$ Jm$^{-3}$ \cite{Kittel-1949}. Its physical density is 9350~kgm$^{-3}$ \cite{Zhang-2011}.
\item Cobalt has $\mu_0M_s=1.81$ T, $K_1=4.1\times10^{5}$ Jm$^{-3}$ \cite{Coey-2010} and $dK_1/dT=-2.1\times10^{5}$ Jm$^{-3}$ \cite{Paige-1984}. Its physical density is 8900~kgm$^{-3}$
\end{itemize}

\noindent Among ferromagnets, MnBi has the unusual property of having $dK_1/dT>0$ at room temperature. This is the consequence of an intrinsic spin reorientation at around $100$ K \cite{Chen-1974}. In MnBi, $K_1$ reaches a maximum at around $500$ K and then decreases. By taking ideal exchange coupling, to obtain zero anisotropy at room temperature, the ratio of layer thicknesses in the proposed multilayer has to be the inverse of the ratio of the anisotropy constants:

\beq
\frac{d_{Co}}{d_{MnBi}} = \frac{K_{1,MnBi}}{K_{1,Co}} \simeq 2.2
\eeq

\noindent The phase fractions in this example will therefore be $p_{Co}=0.69$ and $p_{MnBi}=0.31$. The typical entropy change $\Delta s = dK/dT$ results in $\Delta s = $ 0.34 Jkg$^{-1}$K$^{-1}$. In order to determine the conditions for the exchange coupling between A and B we perfrom a micromagnetic analysis of the multilayer.

\subsection{Micromagnetics of exchange-coupled multilayers}

The problem of the magnetization configuration adopted by a multilayer can be well expressed by micromagnetic theory where the energy of the system is a functional of the magnetization vector field $\mathbf{M}(\mathbf{r})$ and is the sum of several contributions: exchange, anisotropy, magnetostatic and applied field \cite{Bertotti-1998}. The configurations of minimum energy are given by the solution of a variational problem and can be expressed as $\mathbf{m}\times\mathbf{H}_{eff}=0$ in terms of the reduced magnetization $\mathbf{m}=\mathbf{M}/M_s$ and of the effective field, $\mathbf{H}_{eff}$:

\beq
\mathbf{H}_{eff} = \frac{2}{\mu_0 M_s} \nabla \cdot (A\nabla \mathbf{m})-\frac{1}{\mu_0M_s}\frac{\partial f_{AN}}{\partial \mathbf{m}} + \mathbf{H}_{M}+\mathbf{H}_{a}
\eeq

\noindent where the first term on the right hand side is the exchange term and $\nabla \cdot ( A \nabla \mathbf{m} )$ is a shorthand notation for $\nabla \cdot ( A \nabla {m}_x ) \mathbf{i} +\nabla \cdot ( A \nabla {m}_y ) \mathbf{j} +\nabla \cdot ( A \nabla {m}_z ) \mathbf{k} $. The anisotropy energy is $f_{AN}$ and it also contains the presence of a unit vector $\mathbf{n}$, indicating magneto-crystalline orientations.  $\mathbf{H}_{M}$  is the magneto-static field given by the solution of the magneto-static equations ($\nabla\cdot\mathbf{H}_{M} = - \nabla \cdot \mathbf{M}$ and $\nabla\times\mathbf{H}_{M} =0$) and  $\mathbf{H}_{a}$ is the applied magnetic field. In a magnetic multilayer, all parameters change from layer to layer but the exchange term tends to favor the parallel configuration between layers. 
 
The application of micromagnetic theory to exchange coupling in multilayers has been studied in several contexts and for different magnetic materials. Gonzales {\it et al.}\cite{Gonzales-1993} and Navarro {\it et al.} \cite{Navarro-1993} discussed the exchange coupling in hard-soft composites. The problem of the exchange coupling in a magnetic multilayer structure in micromagnetic theory has been studied by Asti {\it et al.} \cite{Asti-2004} who studied the nucleation field of hard-soft multilayer. Alvarez-Prado {\it et al.} \cite{Alvarez-Prado-2007} solved numerically the micromagnetic equation for the case of a bilayer with orthogonal anisotropy axis. Dubuget {\it et al.} \cite{Dubuget-2009} solved numerically the integral equation for the bilayer. Here we solve our problem numerically under rotating magnetic field in order to determine the typical thicknesses which lead to exchange-coupled layers.

\section{Results}
\label{results}

\subsection{Competition between exchange and perpendicular anisotropies}

\begin{figure}
\centering
\includegraphics[width=\columnwidth]{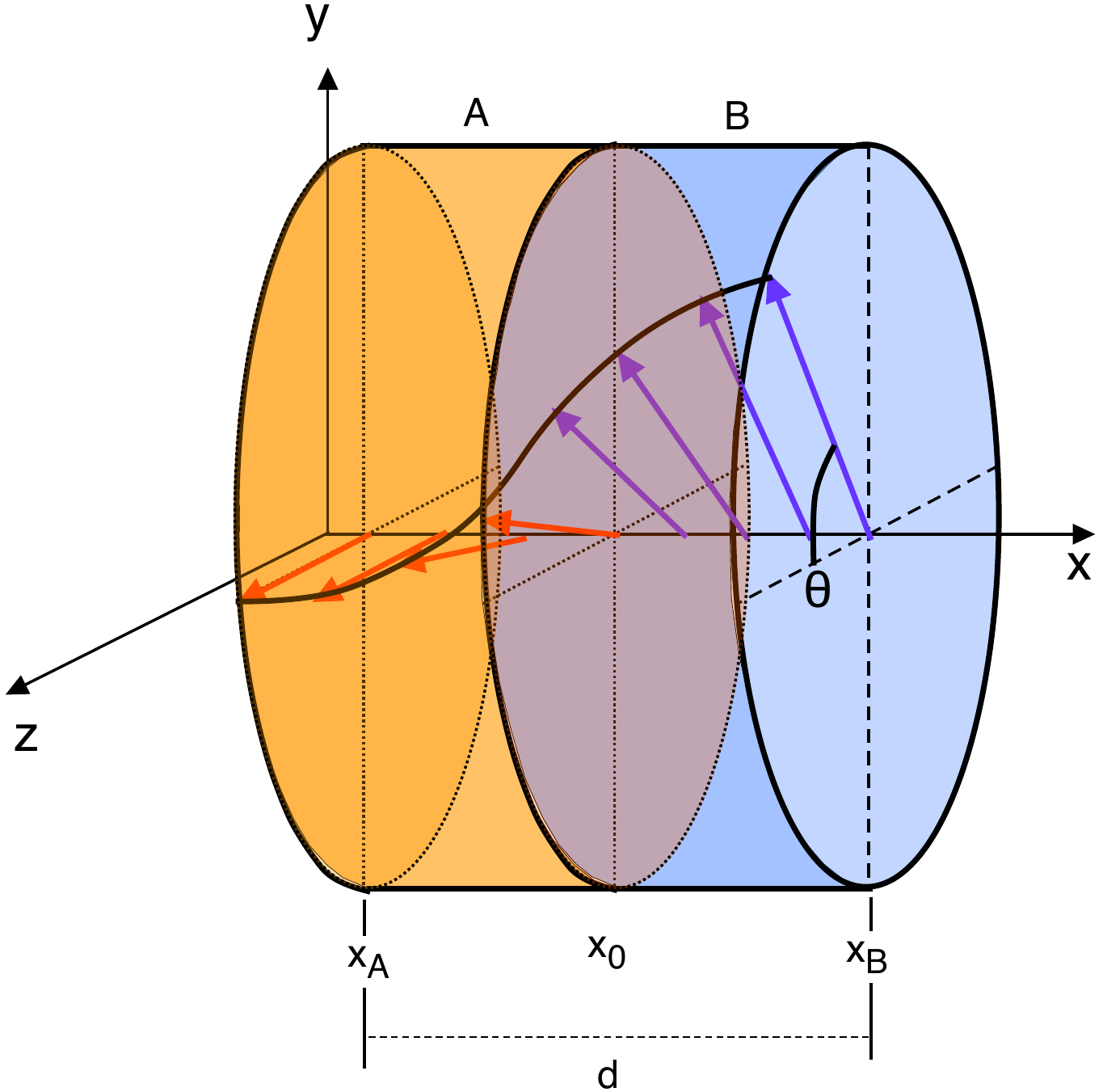}
\caption{Illustration of the angle of the magnetization $\theta(x)$ in the multilayer. The midpoints of layers A and B are at $x_A$ and $x_B$,  respectively.} \label{FIG:1}
\end{figure}

The main feature of the multilayer with perpendicular uniaxial anisotropies is the competition between the exchange, which favors the parallel alignment of different layers, and the local anisotropies, which favor the alignment with the local easy axis. We consider an infinite multilayer composed by an alternating sequence of layers A and B (shown in Fig.~\ref{FIG:1}). We limit to the case in which the magnetic field is in the $(y-z)$ plane and the magnetization changes in space only along $x$. We have $m_x=0$ and $m_y=\sin\theta$, $m_z=\cos\theta$. We let the limits of the system in the $(y-z)$ plane to go to infinity and in this way the configurations are stray field free ($\mathbf{H}_{M}=0$). The thickness of the layer A is $d_A$ and the thickness of the layer B is $d_B$. The sum is $2d=d_A+d_B$. We place the point $x_A$ in the middle of layer A, the point $x_B$ in the middle of layer B and the point $x_0$ at the conjunction of the two layers. Any solution must have a period $2d$ in space and the points $x_A$ and $x_B$ are extrema with $ \left. {\partial \theta}/{\partial x}\right|_{x_A,x_B}=0$. For our case, the only relevant component of the micromagnetic effective field is the one perpendicular to the local magnetization 

\beq
{H}_{eff_{\bot}} = M_s l_{EX}^2 \frac{\partial^2 \theta}{\partial x^2} - H_{AN} \sin\theta\cos\theta + H_{a} \sin(\theta_H-\theta)
\eeq

\noindent where $l_{EX}^2 = {2A}/({\mu_0 M_s^2})$ is the exchange length, $H_{AN} = 2 K /(\mu_0 Ms)$ is the anisotropy field and $\theta_H$ is the angle formed by the applied field with the $z$ axis. In the presence of different layers, the parameters $M_s$, $l_{EX}$ and $H_{AN}$ change values from layer A to layer B and in particular we have positive $H_{AN,A}$  and negative $H_{AN,B}$. The exchange boundary condition at the interface is 

\beq
\left.  A_A \frac{\partial \theta}{\partial x} \right|_{x_0^-} = \left. A_B \frac{\partial \theta}{\partial x}\right|_{x_0^+} \, .
\eeq

\noindent The configurations of minimum energy are given by ${H}_{eff_{\bot}} = 0$. In the following we will derive the configurations realized by a rotating applied field.

\subsection{Rotating applied field}

\begin{figure}
\centering
\includegraphics[width=\columnwidth]{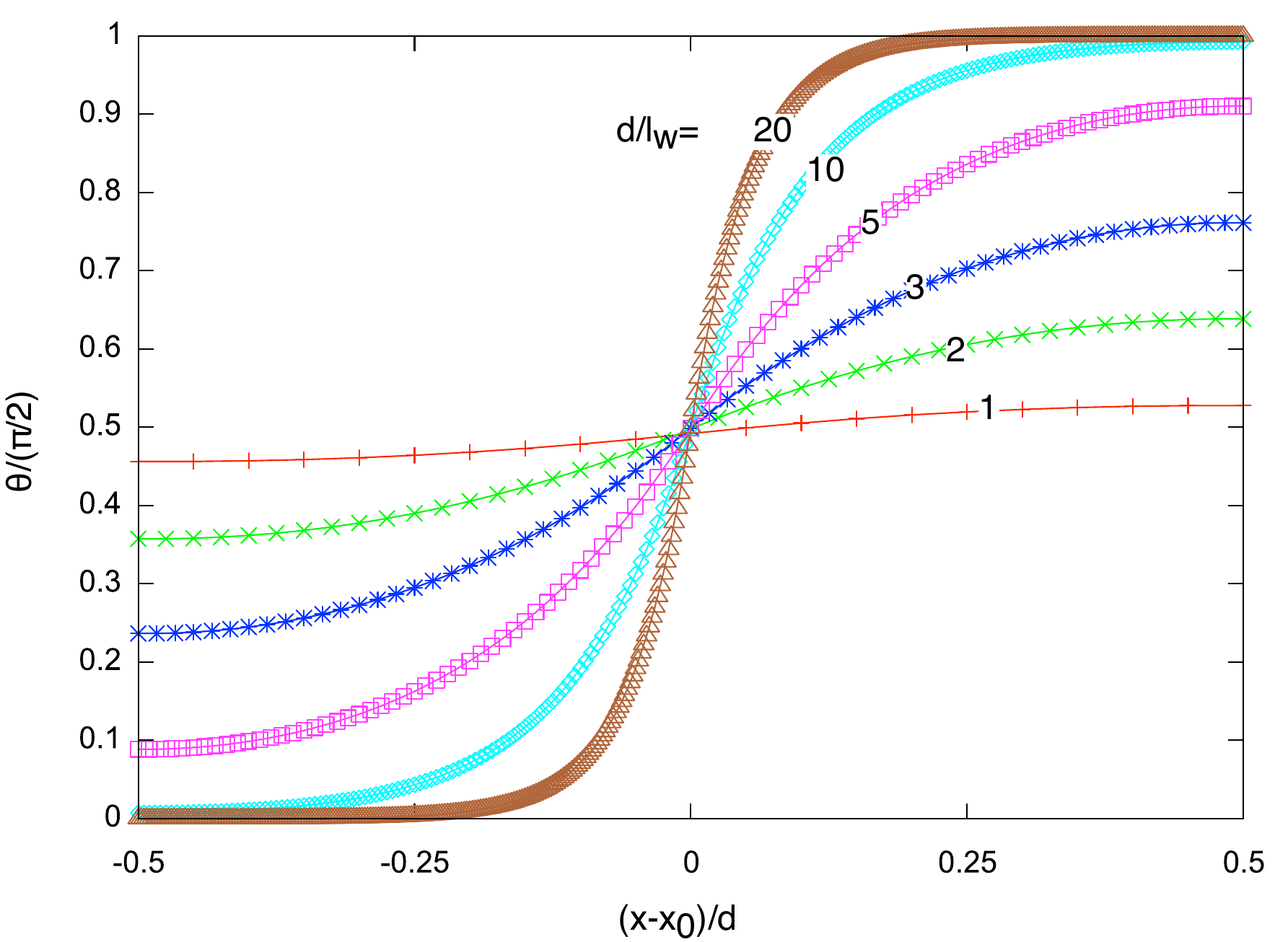}
\caption{Numerical solution of the equilibrium angle $\theta(x)$ between the magnetization and the $z-$ axis (the easy axis of layer A) for several values of the effective layer thickness to domain wall ratio $d/l_w$.  In this case, $h_a=0$.} 
\label{FIG:2}
\end{figure}

\begin{figure}
\centering
\includegraphics[width=\columnwidth]{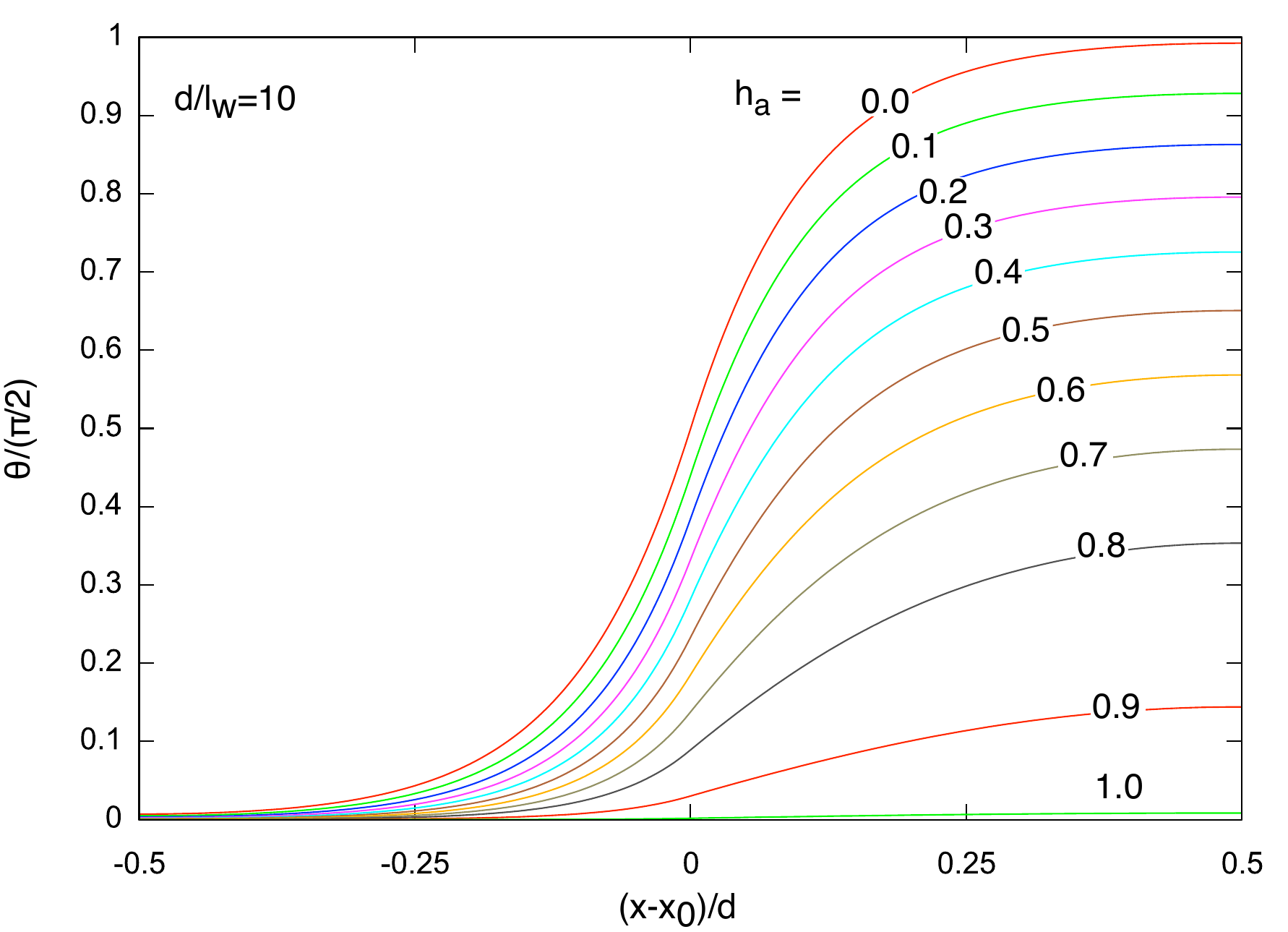}
\caption{Numerical solution of the equilibrium angle $\theta(x)$ for $\theta_H=0$ and several values of $h_a$ with $d/l_w=10$.} 
\label{FIG:3}
\end{figure}

To simplify the problem we select $|H_{AN,A}| = |H_{AN,B}| = H_{AN}$ and normalize all the the fields to $H_{AN}$. We have the normalized effective field

\beq 
h_{eff \bot,i} = l_{w_i}^2 \frac{\partial^2 \theta}{\partial x^2} - \kappa_i \sin\theta\cos\theta + h_{a} \sin(\theta_H-\theta)
\label{EQ: gL}
\eeq

\noindent where $l_{w_i}^2 = {A_i}/{|K_i|}$ is the typical domain wall width for the layer $i = A,B$ and $\kappa_i = \mbox{sgn}(K_i)$. We also choose $A_A = A_B$ and therefore $l_w=l_{w_A}=l_{w_B}$. The configuration of the energy minimum is obtained for each $\theta_H$ by a relaxation method in which the angle $\theta$ is changed according to the equation ${\partial \theta}/{\partial t} = \alpha_G h_{eff,\bot}$, where $\alpha_G$ is the Gilbert damping, until the condition $h_{eff \bot, i} =0$  is met. The solutions were computed numerically with $\kappa_A=1$ and $\kappa_B=-1$ by discretization of the space in $\Delta x/l_w=0.05$ and with time steps $\Delta t \alpha_G=10^{-3}$ and equilibrium was assumed when $|h_{eff,\bot}|<10^{-3}$. Figures~\ref{FIG:2} and \ref{FIG:3} give example outputs for $\theta(x)$. Fig.\ref{FIG:2} shows the angle $\theta$ for several values of $d/l_w$. Fig.~\ref{FIG:3} shows the case $d/l_w=10$ for several values of magnetic field $h_a$ and $\theta_H=0$.

\begin{figure}[htb]
\centering
\includegraphics[width=\columnwidth]{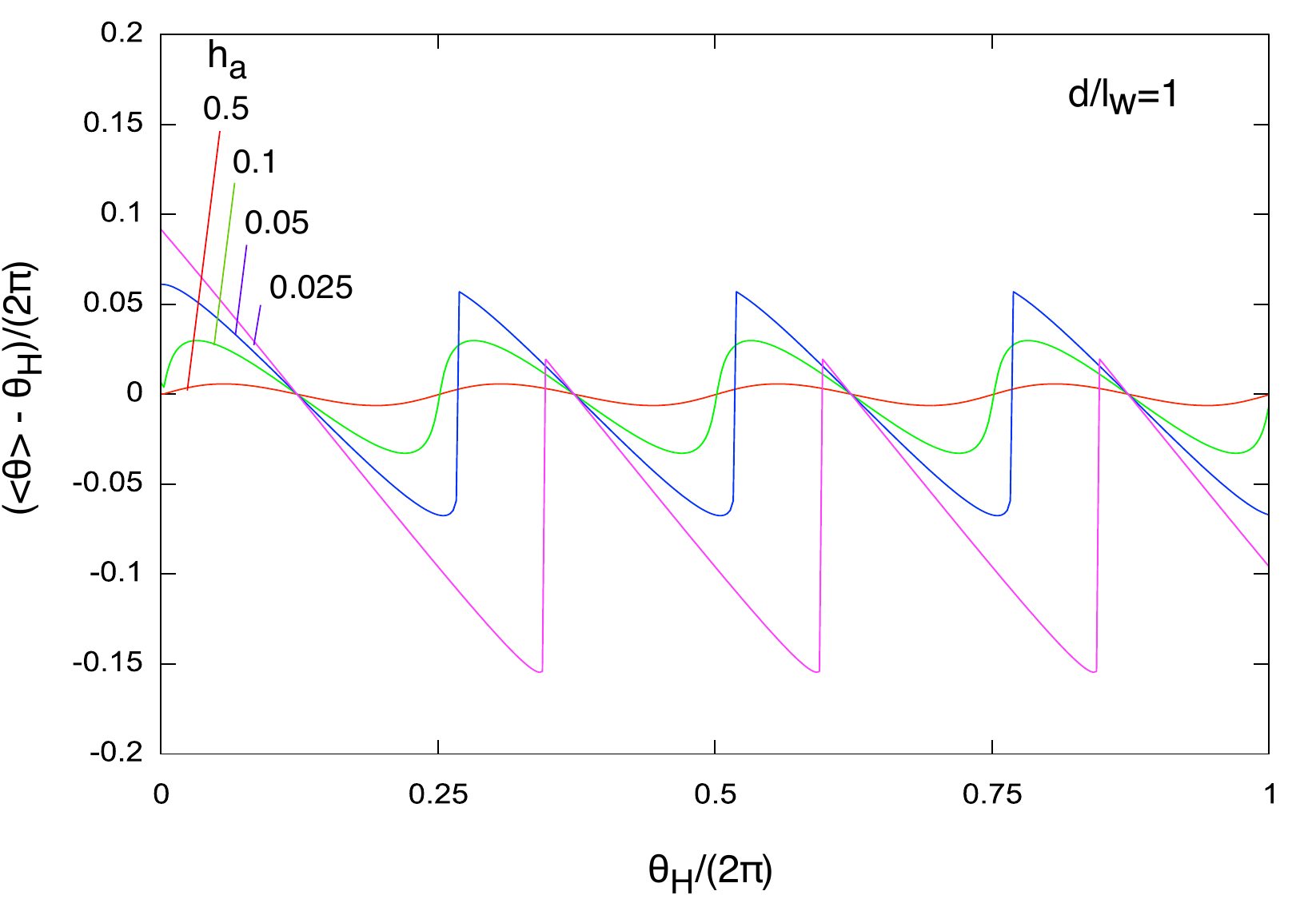}
\caption{Difference between the average magnetization angle $<\theta>$ and magnetic field angle $\theta_H$ for different values of amplitude $h_a$ with $d/l_w=1$} 
\label{FIG:4}
\end{figure}

\begin{figure}[htb]
\centering
\includegraphics[width=\columnwidth]{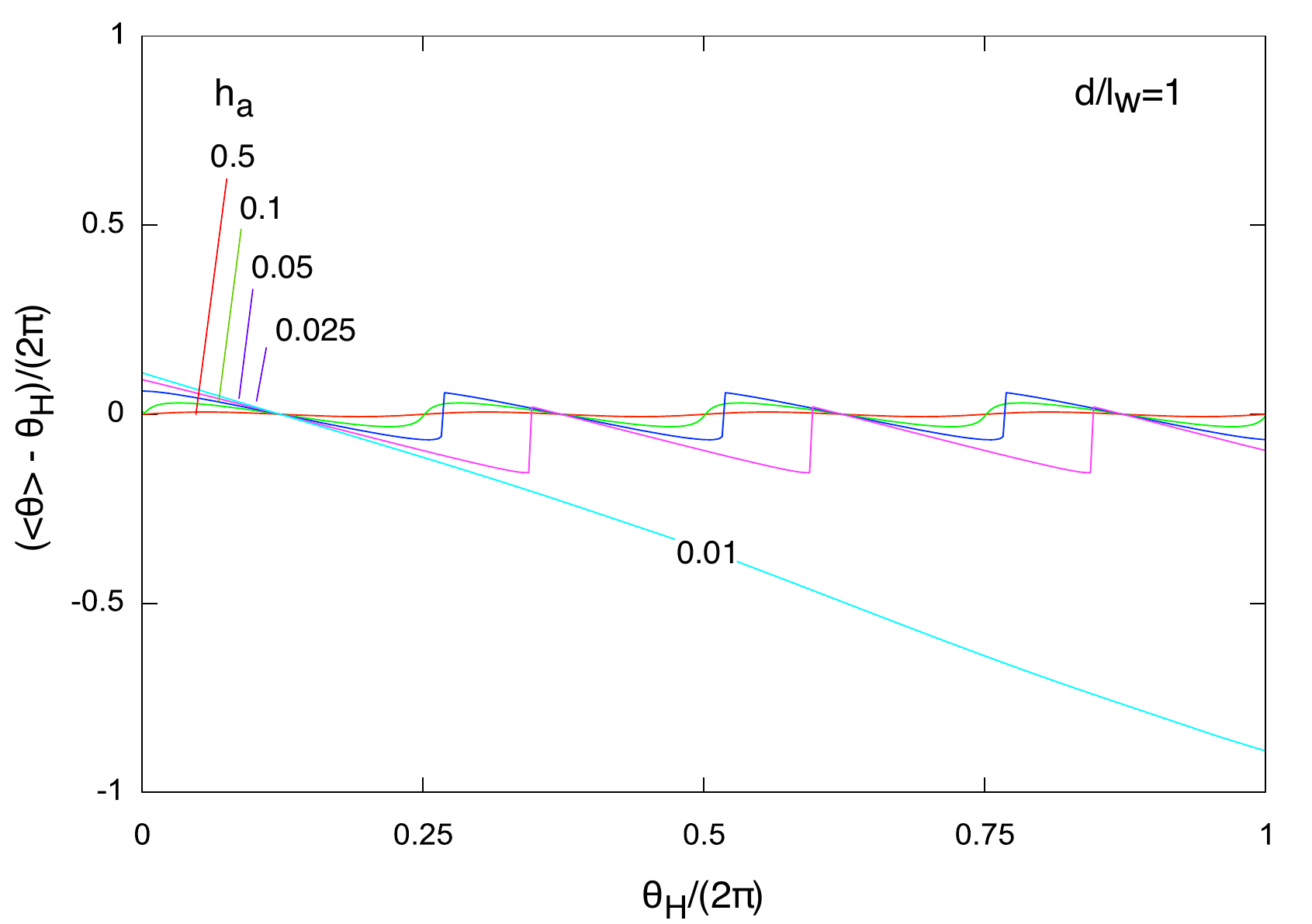}
\caption{Difference between the average magnetization angle $<\theta>$ and magnetic field angle $\theta_H$ for different values of amplitude $h_a$ with $d/l_w=1$. With $h_0=0.01$ the magnetization does not follow the magnetic field for $0.01 < h_0 < 0.025$} \label{FIG:5}
\end{figure}

\begin{figure}[htb]
\centering
\includegraphics[width=\columnwidth]{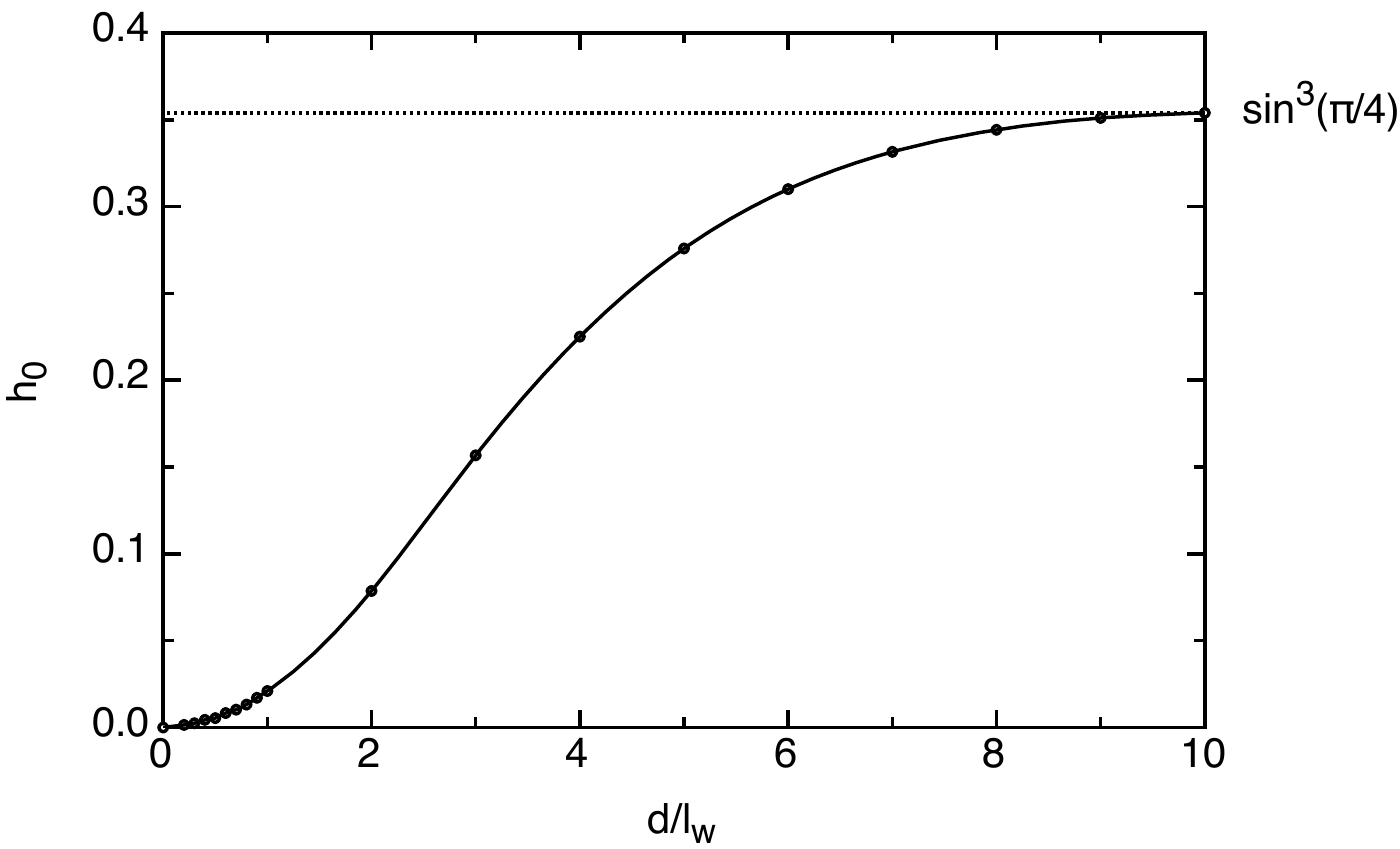}
\caption{Minimum amplitude of rotating magnetic field $h_0$ for which the solution $\theta(x)$ has a compete rotation. In the limit of large $d$ one has the behavior of independent layers with the switching field predicted by the Stoner Wohlfarth model $h_0 = \sin^3(\pi/4)$.  } \label{FIG:6}
\end{figure}

Under rotating magnetic field we are interested in the minimum amplitude of $h_eff=h_0$ for which the full solution $\theta(x)$ has a compete rotation over a distance $2d$ in the $x$-direction.  Fig.~\ref{FIG:4} shows the difference between the average magnetization angle $<\theta>=(1/d) \int \theta(x)dx$ and magnetic field angle $\theta_H$ for different values of amplitude $h_a$ with $d/l_w=1$. In Fig.~\ref{FIG:5} we see that with $h_0=0.01$ the magnetization does not follow the magnetic field. Finally, Fig.~\ref{FIG:6} shows $h_0$ as a function of $d/l_w$.  From Fig.~\ref{FIG:6} one can determine the thicknesses of the layers necessary to achieve the desired coupling. With typical thickness $d \simeq l_w$ the exchange coupling is sufficient to reduce the magnetic field necessary for the full rotation by a factor of 10 with respect to the uncoupled case. 

For a  MnBi/Co multilayer, the average saturation magnetization is $\mu_0<M_s>=1.47$ T and the average anisotropy field is $\mu_0<H_{AN}>=0.95$ T. By using $A=10^{-11}$ Jm$^{-1}$ and $K \simeq 5.6\times10^{5}$ Jm$^{-3}$ as the average amplitude of the anisotropy constants, one has $l_w \sim 4.2$ nm. With the relation $2d=d_{Co}+d_{MnBi}$, we finally obtain $d_{Co}=6.9$ nm and $d_{MnBi}=3.1$ nm. These typical sizes can be achieved by several modern technologies such as sputtering, evaporation, nanoparticles fabrication techniques, etc. With these values the entropy change of the reorientation can be obtained with rotating magnetic field of amplitude $\sim 0.034$ T than can be easily generated within iron yokes by both permanent magnets and electric current windings. 

\section{Conclusions}
\label{conclusions}
In this paper we have computed the magnetocaloric effect associated with the rotation of the magnetization in an exchange coupled multilayer composed of two ferromagnetic materials A and B with uniaxial anisotropy axis perpendicular to each other. Layers A and B are exchange-coupled and their aniostropy constants have  different temperature dependences (positive for A and negative for B).  Such conditions present a spontaneous reorientation of the magnetization at $T_{SR}$ between the easy axes of A and B. We compute the maximum entropy change associated with the reorientation at $T=T_{SR}$. For A=MnBi and B=Co we have an entropy change of $\Delta s = 0.34$ Jkg$^{-1}$K$^{-1}$ for perfect exchange coupling. 

The behavior of the system under rotating magnetic field depends on the ratio between the layer thicknesses and the intrinsic length $l_w = \sqrt{A/K}$. Micromagnetic calculations provide the minimum field amplitude necessary to have a complete rotation of the magnetization as a function of the ratio $d/l_w$ where $d$ is the layer thickness. Our micromagnetic calculations show that for $d \simeq l_w$ the switching field is reduced by a factor 10 with respect to the Stoner Wohlfarth field in the case of no exchange coupling and the maximum entropy change, while small (0.34~Jkg$^{-1}$K$^{-1}$), is reached in a very small magnetic field of only $\sim 0.034$ T.  This represents a $\Delta s/\mu_0 \Delta H$ response of around 10~Jkg$^{-1}$K$^{-1}$T$^{-1}$ which is comparable with that of large magnetocaloric effect materials.  These first results therefore indicate that layered materials with optimized $\frac{dK_1}{dT}$ values should yield yet higher $\Delta s$ and $\Delta s/\mu_0 \Delta H$ values. We finally also note the potential application of such layered materials to thermomagnetic power generation, as described elsewhere~\cite{Tantillo-2023}.

\section*{Acknowledgements}
The research leading to these results has received funding from the European Community’s 7th Framework Programme under Grant Agreement No. 214864 (SSEEC).

\bibliography{srt-bibfile}

\end{document}